\documentclass{article}
\usepackage[letterpaper, scale=0.8]{geometry}
\usepackage{amsmath}
\usepackage{amsfonts}
\usepackage{physics}
\usepackage{graphicx}
\usepackage{subcaption}
\usepackage{float}
\usepackage{booktabs}
\usepackage{multirow}
\usepackage{algorithm}
\usepackage{algorithmic}
\usepackage{cite}
\usepackage{authblk}

\newcommand{\keywords}[1]{%
  \begin{flushleft}
    \textbf{Keywords:} #1
  \end{flushleft}
}

\title{Operator Learning for Reconstructing Flow Fields from Sparse Measurements: a Language Model Approach}
\author[1]{Qian Zhang}
\author[1]{George Em Karniadakis\thanks{george\_karniadakis@brown.edu}}
\affil[1]{Division of Applied Mathematics, Brown University}
\date{}

\begin{document}
\maketitle

\begin{abstract}

  Reconstructing flow fields from sparse measurements is a fundamental problem in fluid mechanics with broad implications for modeling, control, and design. In this work, we propose a novel operator learning framework that leverages the architecture of language models to perform flow reconstruction in a mesh-free manner. We reformulate flow field reconstruction as a sequence-to-sequence learning task, where sparse measurements are treated as context and unobserved locations as queries. Our model learns to reconstruct the full flow field from sparse inputs, effectively capturing spatial correlations and long-range dependencies. We evaluate the proposed approach on four benchmark datasets: (1) two-dimensional vortex street simulations, (2) daily average temperature data across the contiguous United States, (3) three-dimensional blood flow simulations based on dissipative particle dynamics, and (4) three-dimensional turbulent jet flow measurements obtained via particle tracking velocimetry. Across all cases, our method demonstrates competitive reconstruction accuracy, even with highly incomplete data (less than 10\% observed), and achieves efficient performance. The results highlight the potential of language models as robust and scalable tools for scientific data reconstruction, and suggest a promising direction toward the development of foundation models for scientific and engineering applications.

\end{abstract}

% keywords

\section{Introduction}

Recent advances in machine learning have led to significant breakthroughs across a wide range of scientific and engineering disciplines~\cite{price2024weather,AlphaTensor2022,cai2022review,carleo2017quantum,jin2020SympNets,zhang2024sms,zhang2025artificial,theilman2024loihi,wu2023comprehensive,wan2024rfg,jumper2021protein,merchant2023materials,cao2024lno,daneker2023sysbio,zhang2022AOSLO,zhang2021PLOS,kharazmi2021}. Problems previously considered intractable by conventional approaches can now be reformulated as optimization tasks solvable by neural network models, benefiting from the rapid growth of computational resources.

Within this paradigm, operator learning has emerged as a powerful framework for modeling complex systems~\cite{lu2021learning,cao2024lno,goswami2022physics,kahana2022spiking,zhang2024hints}. Operator learning aims to use neural network to approximate solution operators, which are mappings from input functions or initial/boundary conditions to output solutions. These neural operators serve as efficient surrogates to traditional numerical solvers, offering significantly faster inference, especially in high-dimensional or real-time scenarios.

Several recent developments have demonstrated the capabilities of neural operators in different aspects. DeepONet~\cite{lu2021learning} provides a mesh-free framework that enables querying at arbitrary points in the domain. The Fourier Neural Operator (FNO)~\cite{li2020fourier} uses Fourier transforms to model long-range dependencies and global structures in spatial-temporal data. The Laplace Neural Operator (LNO)~\cite{cao2024lno} leverages Laplace transforms to handle transient and non-periodic dynamics. ViTO~\cite{ovadia2024vito}, based on Vision Transformers, demonstrates efficient operator learning on regular meshes by incorporating transformer-based architectures into the operator learning pipeline.

Despite this progress, applying operator learning to reconstruction tasks remains challenging, especially in settings where the observations are sparse, noisy, and irregularly distributed. A representative example is jet flow reconstruction in fluid mechanics, where only a small portion of the flow field is accessible through measurements, often not aligned on a structured grid.

A number of approaches have been proposed for sparse reconstruction and super-resolution in scientific contexts. Data-driven sparse sensor placement methods~\cite{manohar2018sparse} optimize observation locations to maximize reconstruction quality. Shallow decoder networks~\cite{erichson2020shallow} use limited sensors combined with a linear basis to reconstruct full fields. Voronoi-tessellation-assisted CNNs~\cite{fukami2021voronoi} handle irregularly spaced sensor data by constructing Voronoi diagrams as a preprocessing step. On the architectural side, set-based architectures such as the Set Transformer~\cite{lee2019set} and Perceiver IO~\cite{jaegle2021perceiver} provide permutation-invariant processing of unordered input sets, while point-cloud transformers~\cite{zhao2021point} operate directly on unstructured 3D point data. The Physics-Informed Diffusion Model (PIDM)~\cite{shu2023pidm} exhibits promise in reconstructing flow fields from sparse inputs, but requires measurements on a fixed, regular mesh. Our prior work~\cite{zhang2025operator} extended operator learning to reconstruction settings, but still relied on structured data, which constrains its applications to more realistic and irregular scenarios.

Recent developments in large language models (LLMs)~\cite{openai2024gpt4,deepseek2025r1} have demonstrated remarkable ability to model complex patterns and long-range dependencies in the data. In parallel, the concept of in-context learning — where a model learns to perform a task from examples provided in its input, without weight updates — has been explored for scientific machine learning~\cite{yang2023context,yang2025fine,yang2024pde,cao2024vicon}. This in-context paradigm aligns naturally with the Conditional Neural Process (CNP) family~\cite{garnelo2018conditional,kim2019attentive}, which formalizes the mapping from a context set of observations to predictions at query locations. Inspired by these advancements, we propose RFormer (Reconstruction Former), an in-context operator learning framework that reformulates the sparse reconstruction problem using the architecture of decoder-only transformers. More specifically, we interpret observed data points as context tokens and unobserved locations as query tokens. The goal is to predict the missing values, analogous to the answer tokens in a question-answering (Q\&A) framework. This formulation naturally aligns with transformer architectures and enables the model to generalize to arbitrary query locations without requiring structured meshes.

In this paper, we present a novel operator learning framework that integrates transformer-based language models to perform mesh-free field reconstruction. We demonstrate the efficacy of our method on four diverse and challenging datasets, including real-world scenarios with noisy measurements, thereby highlighting its generality and robustness in practical scientific applications.

\section{Method}
\subsection{Related work}
\textbf{Traditional methods.} Classical sparse-reconstruction methods provide strong and interpretable baselines for this problem. Interpolation methods~\cite{lee1980delaunay} estimate unknown values directly from nearby observations and are simple to apply on irregular point sets, but they rely mainly on local smoothness and can struggle with coherent structures far from sensors. Kriging~\cite{oliver1990kriging} extends interpolation with a probabilistic spatial covariance model, making it a principled statistical approach when the covariance assumptions are appropriate. Gappy POD~\cite{everson1995karhunen} reconstructs missing fields using a low-dimensional basis learned from training snapshots; it can be highly effective when test states remain close to the training trajectory but is less flexible under changes in geometry, dynamics, or sampling pattern.

\textbf{Voronoi CNN.} Voronoi-tessellation CNNs provide a neural alternative for sparse flow reconstruction from irregular sensors~\cite{fukami2021voronoi}. They convert scattered sensor measurements into a grid-like representation by assigning regions of the domain to nearby sensors through a Voronoi tessellation, after which standard convolutional networks can be applied. This approach is effective for two-dimensional fields where the irregular observations can be rasterized onto an image-like domain, but the preprocessing and convolutional architecture are less natural for fully unstructured three-dimensional point clouds.

\textbf{Neural processes.} The neural process family is another closely related line of work because it also formulates prediction as a mapping from context observations to query locations. Conditional Neural Processes (CNPs)~\cite{garnelo2018conditional} aggregate context information into a global representation before decoding query predictions, Attentive Neural Processes (ANPs)~\cite{kim2019attentive} add attention to improve query-specific conditioning, and Transformer Neural Processes (TNPs)~\cite{nguyen2022transformer} use transformer-style attention to model richer interactions among context and query points. These methods are natural baselines for our setting, but our formulation is tailored to operator-learning reconstruction with explicit position-value tokens, a structured attention mask, and chunked inference for large scientific point sets.

\subsection{Operator learning framework}
In this section, we formulate an operator learning framework for the reconstruction problem and introduce the proper notation. These are the same as in our previous work~\cite{zhang2025operator}, but we point out that it can handle irregular data positions and is more general than the applications in our previous study.

We denote a data sample as a collection of position-value pairs $\mathcal{P}_a=\{(x_a^A, v_a^A)\}_{A=1}^{M_a}$, where $x_a^A$ represents the position, $v_a^A$ denotes the data value at position $x_a^A$, and $M_a$ is the number of position-value pairs in sample $a$. This collection $\mathcal{P}_a$ is referred to as a \textbf{full} data sample. Meanwhile, an \textbf{observed} data sample is defined as $\tilde{\mathcal{P}}_a=\{(x_a^A, v_a^A)\}_{A\in\mathcal{O}_a}$, where $\mathcal{O}_a$ indexes the observed position-value pairs. A \textbf{dataset} is defined as a collection of full data samples $\mathcal{D}=\{\mathcal{P}_a\}_{a=1}^{M}$, where $M$ is the number of samples. The goal of operator learning for the reconstruction problem is to learn a mapping $\mathcal{R}$ from the observed data sample to the full data sample, i.e.,
\begin{equation}
  v_a^{A'_j} = \mathcal{R}(x_a^{A'_j}; \tilde{\mathcal{P}}_a), \quad \forall A'_j\not\in\mathcal{O}_a, \; j=1,\ldots,n.
\end{equation}

In the language of operator learning, the learned mapping $\mathcal{R}$ is an operator that maps from the space of sparsely observed functions (defined only on the subset of positions indexed by $\mathcal{O}_a$) to the space of fully reconstructed functions (defined at all positions). Since the model accepts arbitrary spatial coordinates $x_a^{A'_j}$ as input rather than being tied to a fixed discretization grid, the framework is discretization-invariant — a defining characteristic of neural operators in the sense of Kovachki and co-workers \cite{lu2021learning,li2020fourier}. This positions our method as a natural extension of operator learning to sparse reconstruction settings on unstructured data.

Now we interpret the formula into the language model framework. We use observed data sample as the context, and the unknown data positions as the questions, true values at unknown data positions as the answers.
\begin{equation}
  \underbrace{(x_a^{A_1}, v_a^{A_1}), \ldots, (x_a^{A_m}, v_a^{A_m})}_{\text{context tokens}}, \underbrace{x_a^{A'_1}, \ldots, x_a^{A'_n}}_{\text{query tokens}} \rightarrow \underbrace{v_a^{A'_1}, \ldots, v_a^{A'_n}}_{\text{answer tokens}},
  \label{eq:framework}
\end{equation}
where $A_1, \ldots, A_m$ are the indices of observed data positions (i.e., $A_i \in \mathcal{O}_a$), and $A'_1, \ldots, A'_n$ are the indices of unknown data positions (i.e., $A'_j \notin \mathcal{O}_a$).

This context-query view is most closely related to in-context learning: the observed position-value pairs act as the context supplied at inference time, and the query positions specify the missing values to be inferred without any task-specific parameter update. It is also related to the neural process family, especially ANP and TNP, which condition predictions at query locations on observed context pairs. However, RFormer is not a simple extension of ANP or TNP. Instead of using the standard neural-process encoder-decoder abstraction, we cast reconstruction as a decoder-only token prediction problem with explicit position-value tokens, interleaved observation and query tokens, and a structured attention mask designed for scientific field reconstruction. The target is also an operator-learning reconstruction map over large unstructured point sets, with chunked inference used to assemble full fields, rather than a generic conditional stochastic-process regression model.

\subsection{Network architecture and training strategy}
\begin{figure}[H]
  \centering
  \includegraphics[width=0.6\textwidth]{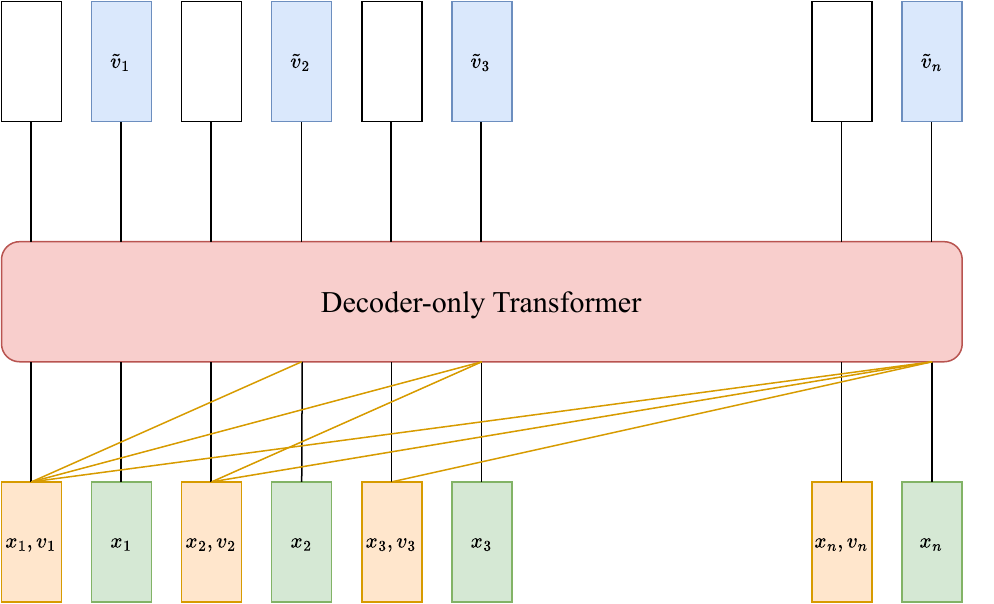}
  \caption{The architecture of the reconstruction framework. Observation tokens (orange, carrying both position and value) and query tokens (green, carrying position only, with values set to zero) are interleaved to form the input sequence. The decoder-only transformer processes them through stacked self-attention blocks with a custom attention mask. Outputs at query-token positions (blue) are used to compute the loss; outputs at observation positions are discarded.}
  \label{fig:method:framework}
\end{figure}

To implement the framework, we adopt a decoder-only transformer architecture~\cite{vaswani2017attention,radford2019language} with several design choices tailored to the sparse reconstruction setting. The overall framework is visualized in Figure~\ref{fig:method:framework}.

\paragraph{Token construction and embedding.}
Unlike natural language processing where tokens are drawn from a discrete vocabulary, our tokens are continuous vectors encoding physical quantities. Each observation point is represented as a concatenated vector $[x, v] \in \mathbb{R}^{d_x + d_v}$, where $x$ is the spatial coordinate and $v$ is the field value at that location. Each query point is represented as $[x, \mathbf{0}] \in \mathbb{R}^{d_x + d_v}$, where the value channel is filled with zeros to indicate its unknown status. These raw vectors are embedded into a $d_{\text{token}}$-dimensional latent space via a single learned linear projection. No positional encoding is used, as the spatial coordinates are already carried explicitly in the token representation — the model learns spatial relationships directly from the coordinate-valued channels.

Given $m$ observation points and $n$ query points, the input sequence alternates between observation and query tokens:
\begin{equation}
  \underbrace{[x_1, v_1]}_{\text{obs token }1},\,
  \underbrace{[x'_1, \mathbf{0}]}_{\text{query token }1},\,
  \underbrace{[x_2, v_2]}_{\text{obs token }2},\,
  \underbrace{[x'_2, \mathbf{0}]}_{\text{query token }2},\,
  \ldots,\,
  \underbrace{[x_m, v_m]}_{\text{obs token }m},\,
  \underbrace{[x'_n, \mathbf{0}]}_{\text{query token }n}.
\end{equation}
This notation is schematic: the interleaved format is used to define the attention pattern and does not require paired observation and query tokens to be spatial neighbors. Padding is only needed when the final evaluation query chunk has fewer than $n$ query points, as described below.

\paragraph{Transformer backbone and attention mask.}
The embedded sequence is processed by a stack of $L$ identical self-attention blocks, each comprising multi-head self-attention~\cite{vaswani2017attention} with a residual connection followed by layer normalization, and a feed-forward network with GELU activation and a second residual connection and layer normalization.

Since the input sequence has no natural causal order, we do \emph{not} apply a standard causal (autoregressive) mask. Instead, we design a custom attention mask that encodes the known structure of the reconstruction problem: query tokens are allowed to attend to all observation tokens (which carry known values) but not to other query tokens (which carry only position information with zero-filled values). Observation tokens may attend to all previous observation tokens and themselves. This structured masking ensures that query tokens aggregate information exclusively from the observed data, preventing contamination from uninformative query-to-query interactions.

\paragraph{Output decoding and loss function.}
The transformer output at each query-token position is decoded through a two-layer MLP with GELU activation to produce the predicted field value $\hat{v} \in \mathbb{R}^{d_v}$. Outputs at observation-token positions are discarded. The training objective is the relative root mean square error (Relative RMSE), computed only over query positions:
\begin{equation}
  \mathcal{L} = \frac{\sqrt{\sum_{i=1}^{n} \|\hat{v}_i - v_i\|^2}}{\sqrt{\sum_{i=1}^{n} \|v_i\|^2}},
\end{equation}
where $v_i$ and $\hat{v}_i$ are the true and predicted values at query position $i$. This normalized loss is scale-invariant, making it suitable across datasets with different physical units and magnitudes.

\paragraph{Batching and padding strategy.}
Each dataset sample may contain many more spatial points than can be processed in one transformer pass. During training, we therefore sample a fixed number of observations and query targets from a single snapshot at every step. This stochastic subsampling exposes the model to different context and query sets across epochs while keeping the sequence length bounded. During evaluation, one observation subset is sampled for each snapshot and reused while the query points are processed in chunks. The chunk predictions are then assembled back into the original point ordering to recover the full field. If the final query chunk is shorter than the prescribed query count, it is padded and the padded entries are excluded from the loss and from the reconstructed output.

\paragraph{Model size and efficiency.}
The architecture is intentionally compact, using the same hyperparameters across all experiments as summarized in Table~\ref{tab:experiments:hyperparameters}. Its parameter count is approximately 350,000 and does not scale with the number of spatial points in a snapshot. Large fields are handled by query chunking at inference time, so memory usage is controlled by the per-pass sequence length rather than by the full resolution of the target field. This makes the method practical for the largest case considered here, where a full blood-flow snapshot contains roughly 300,000 points.

\section{Numerical Experiments}
\label{sec:experiments}
We present four numerical experiments to demonstrate the effectiveness of the proposed method. Both synthetic and real datasets are used, and the results indicate that the proposed method achieves competitive performance with only a small amount of observed data. We begin with the synthetic vortex-street dataset, where controlled ablations over observation density and measurement noise can be performed, and then proceed to real weather, blood-flow, and turbulent-jet datasets. Moreover, the method is computationally efficient and does not require intensive hyperparameter tuning. Table~\ref{tab:experiments:hyperparameters} summarizes the training setup for all experiments.

\begin{table}[H]
  \centering
  \begin{tabular}{ccc}
    \toprule
    Type                   & Hyperparameter   & Value         \\
    \midrule
    \multirow{4}{*}{Model} & Number of layers & 4             \\
                           & Number of heads  & 8             \\
                           & Token size       & 128           \\
                           & Activation       & GELU          \\
    \midrule
    \multirow{6}{*}{Training}
                           & Loss function    & Relative RMSE \\
                           & Optimizer        & Adam          \\
                           & Learning rate    & $0.001$       \\
                           & Batch size       & 16            \\
                           & Sequence length  & $\leq1024$    \\

    \bottomrule
  \end{tabular}
  \caption{Model hyperparameters (the same for all four examples).}
  \label{tab:experiments:hyperparameters}
\end{table}

We compare RFormer against several reconstruction baselines: linear interpolation, Kriging, Gappy POD, Conditional Neural Processes (CNP), Attentive Neural Processes (ANP), Transformer Neural Processes (TNP), and the Voronoi-tessellation CNN baseline. The evaluation metric used is the relative root mean square error (Relative RMSE), defined as
\begin{equation}
  \text{Relative RMSE} = \frac{\sqrt{\sum_{i=1}^{N} (v_i - \hat{v}_i)^2}}{\sqrt{\sum_{i=1}^{N} v_i^2}},
\end{equation}
where $v_i$ is the true value, $\hat{v}_i$ is the predicted value, and $N$ is the number of data points. The configurations of the baselines are detailed in the Appendix.

\subsection{2D Vortex Street}
Vortex streets are repeating patterns of alternating vortices formed by the unsteady separation of flow around bluff bodies, such as cylinders. They are a fundamental phenomenon in fluid dynamics and serve as canonical examples of flow instability and transition to turbulence. Understanding and modeling vortex street dynamics are critical not only for validating computational fluid dynamics (CFD) solvers, but also for practical applications such as flow-induced vibration prediction, noise reduction, and active flow control.

In this experiment, we evaluate our method on the reconstruction of a two-dimensional vortex street dataset generated by simulating incompressible flow past a circular cylinder. The flow is governed by the incompressible Navier–Stokes equations and numerically solved using a high-fidelity spectral element method. The resulting dataset consists of 100 temporal snapshots, each containing approximately 29,000 spatial data points that describe the horizontal velocity ($u$), vertical velocity ($v$), and pressure ($p$) fields.

\begin{figure}[H]
  \centering
  \includegraphics[width=0.8\linewidth]{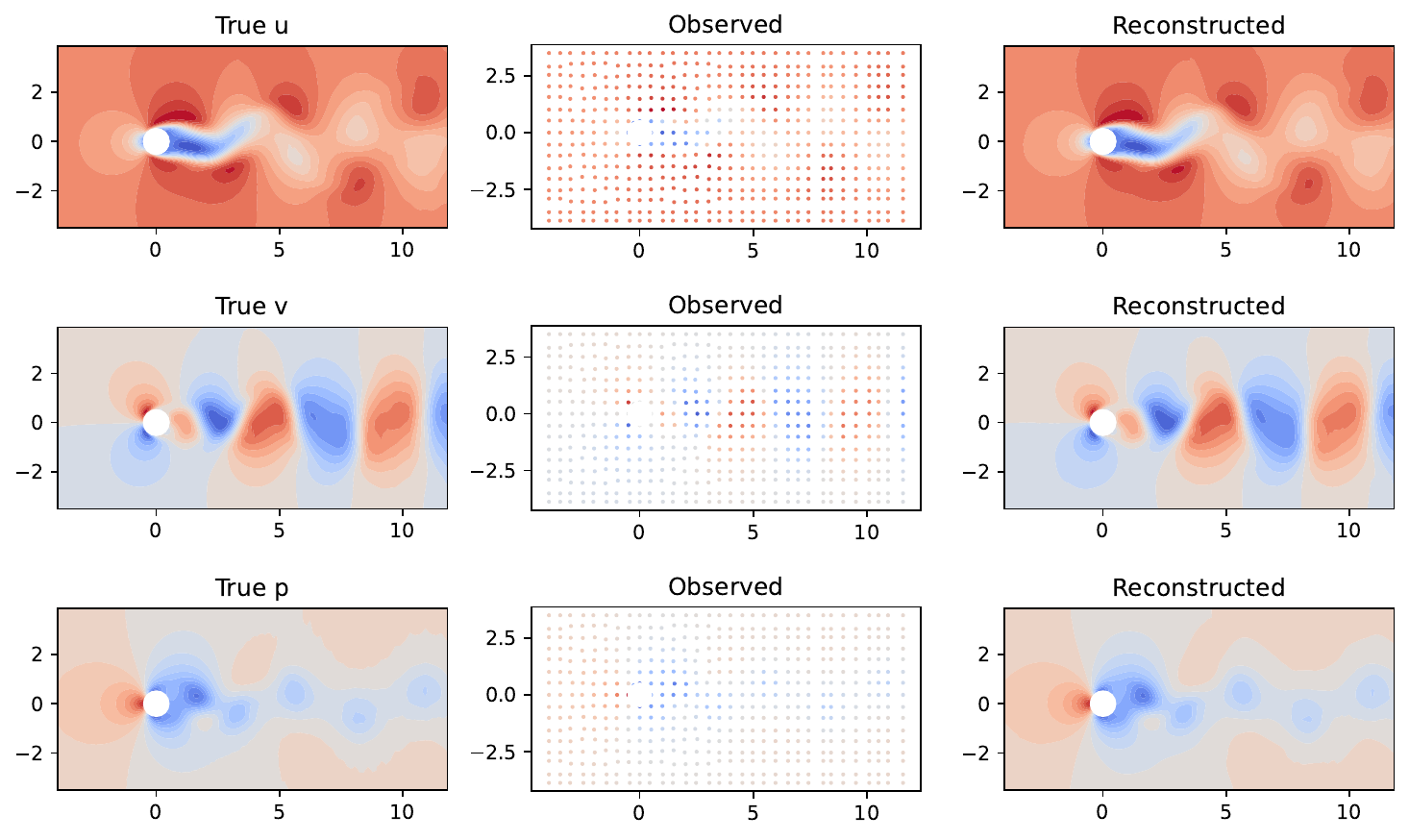}
  \caption{True flow field (left), observed flow field (middle), and reconstructed field (right) for snapshot index 90. The first row shows horizontal velocity $u$, the second row vertical velocity $v$, and the third row pressure $p$. The model accurately captures the flow features despite the highly sparse observations.}
  \label{fig:experiments:vortex:snapshots}
\end{figure}

To emulate a realistic sparse sensing scenario, we sample approximately 500 observation points per snapshot to achieve uniform coverage and treat the remaining points as unknowns to be reconstructed. The first 80 snapshots (indices 0--79) are used for training, and the remaining 20 snapshots (indices 80--99) are reserved for testing. We note that this sequential temporal split means train and test snapshots come from the same simulation trajectory and are therefore temporally adjacent; the reconstruction task thus tests generalization to a later time window of the same vortex-shedding regime rather than to an entirely independent flow configuration. The model is trained for 100 epochs and the training process takes approximately three hours on a single NVIDIA A6000 GPU. The reconstruction results are shown in Figure~\ref{fig:experiments:vortex:snapshots}.

\begin{table}[H]
  \centering
  \begin{tabular}{lcccc}
    \toprule
    Method        & $u$                      & $v$                      & $p$                      & Total                    \\
    \midrule
    RFormer       & $0.0061 \pm 0.0008$          & $0.0204 \pm 0.0047$          & $0.0101 \pm 0.0021$          & $\mathbf{0.0080 \pm 0.0014}$ \\
    Voronoi CNN   & $0.0138 \pm 0.0008$          & $0.0465 \pm 0.0025$          & $0.0244 \pm 0.0008$          & $0.0184 \pm 0.0008$          \\
    Gappy POD     & $0.0197 \pm 0.0018$          & $0.1082 \pm 0.0082$          & $0.0408 \pm 0.0046$          & $0.0338 \pm 0.0028$          \\
    CNP           & $0.0873 \pm 0.0031$          & $0.5606 \pm 0.0222$          & $0.1349 \pm 0.0090$          & $0.1620 \pm 0.0056$          \\
    ANP           & $0.0868 \pm 0.0034$          & $0.5879 \pm 0.0167$          & $0.1810 \pm 0.0299$          & $0.1709 \pm 0.0046$          \\
    Interpolation & $0.1472 \pm 0.0006$          & $0.4214 \pm 0.0018$          & $0.2484 \pm 0.0004$          & $0.1843 \pm 0.0004$          \\
    TNP           & $0.1971 \pm 0.0019$          & $0.8855 \pm 0.0469$          & $0.4448 \pm 0.0296$          & $0.3082 \pm 0.0100$          \\
    KRIGING       & $0.5463 \pm 0.0009$          & $0.9968 \pm 0.0017$          & $0.9621 \pm 0.0061$          & $0.6243 \pm 0.0012$          \\
    \bottomrule
  \end{tabular}
  \caption{Baseline comparison for the vortex-street case. Values are test-set Relative RMSE reported as mean $\pm$ standard deviation across snapshots; lower is better.}
  \label{tab:experiments:vortex_street}
\end{table}

As shown in Table~\ref{tab:experiments:vortex_street}, RFormer gives the lowest error across all three physical variables, with a total Relative RMSE of $0.0080$. Voronoi CNN is the second-best method with total error $0.0184$, followed by Gappy POD at $0.0338$. The neural-process baselines are less accurate on this case: CNP and ANP have comparable total errors of $0.1620$ and $0.1709$, while TNP reaches $0.3082$. Linear interpolation performs better than TNP but remains well above the learned and POD-based reconstructions, and Kriging has the largest error among the tested methods.

\begin{figure}[H]
  \centering
  \begin{subfigure}{0.49\textwidth}
    \includegraphics[width=\textwidth]{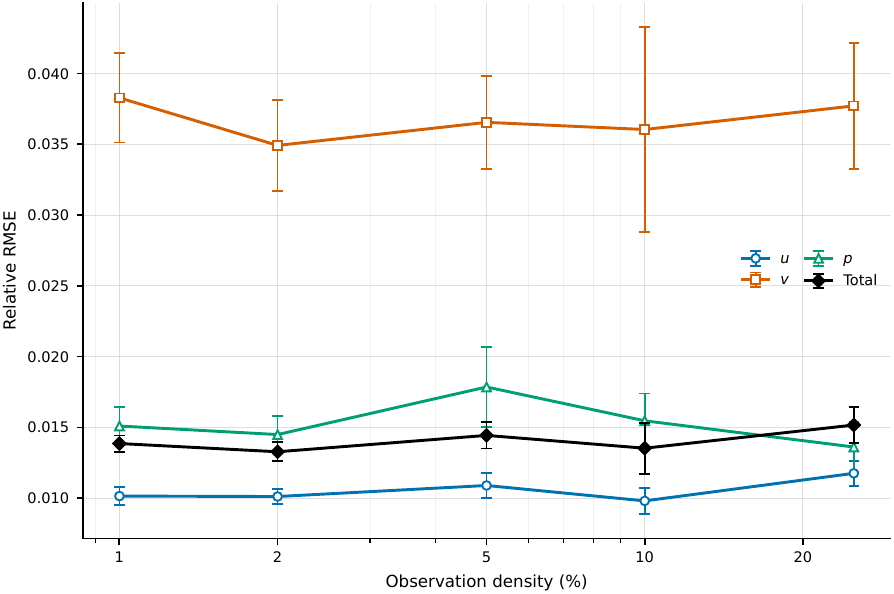}
    \caption{Observation-density ablation}
  \end{subfigure}
  \begin{subfigure}{0.49\textwidth}
    \includegraphics[width=\textwidth]{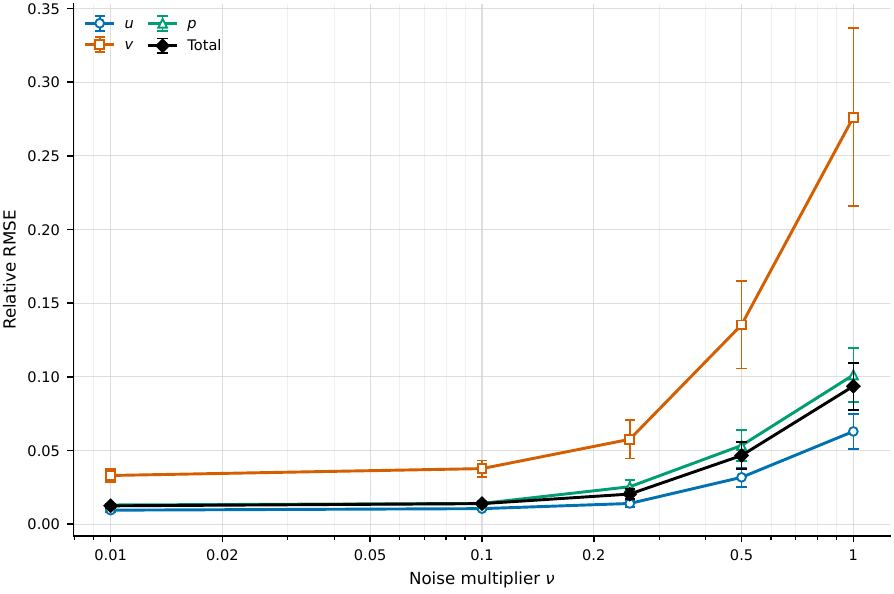}
    \caption{Measurement-noise ablation}
  \end{subfigure}
  \caption{Ablation study on the vortex-street dataset. Error bars denote one standard deviation across test snapshots. The density ablation shows that the model remains accurate even with very sparse observations, with total Relative RMSE between 0.0133 and 0.0152 across 1\%--25\% observation density. The noise ablation shows graceful degradation as observation noise increases: total Relative RMSE rises from 0.0125 at $0.01\times$ noise to 0.0937 at $1.0\times$ noise.}
  \label{fig:experiments:vortex:ablation}
\end{figure}

To assess robustness to sparse and noisy measurements, we further conduct two ablation studies on the vortex-street dataset while keeping the model architecture and training budget fixed. First, we vary the observation density from approximately 1\% to 25\% of the spatial points. Second, we add zero-mean Gaussian noise to the observed field values, with standard deviation ranging from $0.01$ to $1.0$ times the per-component standard deviation of the training data. Figure~\ref{fig:experiments:vortex:ablation} summarizes the resulting test-set Relative RMSE.

The density ablation indicates that the method does not depend strongly on the exact number of observation points within the tested range. Total Relative RMSE is 0.0139 at approximately 1\% observation density and remains comparable at higher densities, with component-wise errors consistently lowest for $u$ and highest for $v$. The noise ablation shows the expected monotonic degradation: total Relative RMSE is 0.0140 at $0.1\times$ noise, 0.0205 at $0.25\times$ noise, 0.0467 at $0.5\times$ noise, and 0.0937 at $1.0\times$ noise. These results indicate that the reconstruction is robust to substantial sparsity and moderate measurement noise, while high noise levels remain challenging.

\subsection{Contiguous U.S. Weather}
To evaluate the performance of our method in a real-world problem, we use the daily average temperature data from the contiguous United States, provided by the National Oceanic and Atmospheric Administration (NOAA). We consider data from the years 2020 to 2024 and select weather stations with complete records over this period, resulting in a total of 2,020 stations.

\begin{figure}[H]
  \centering
  \includegraphics[width=0.6\textwidth]{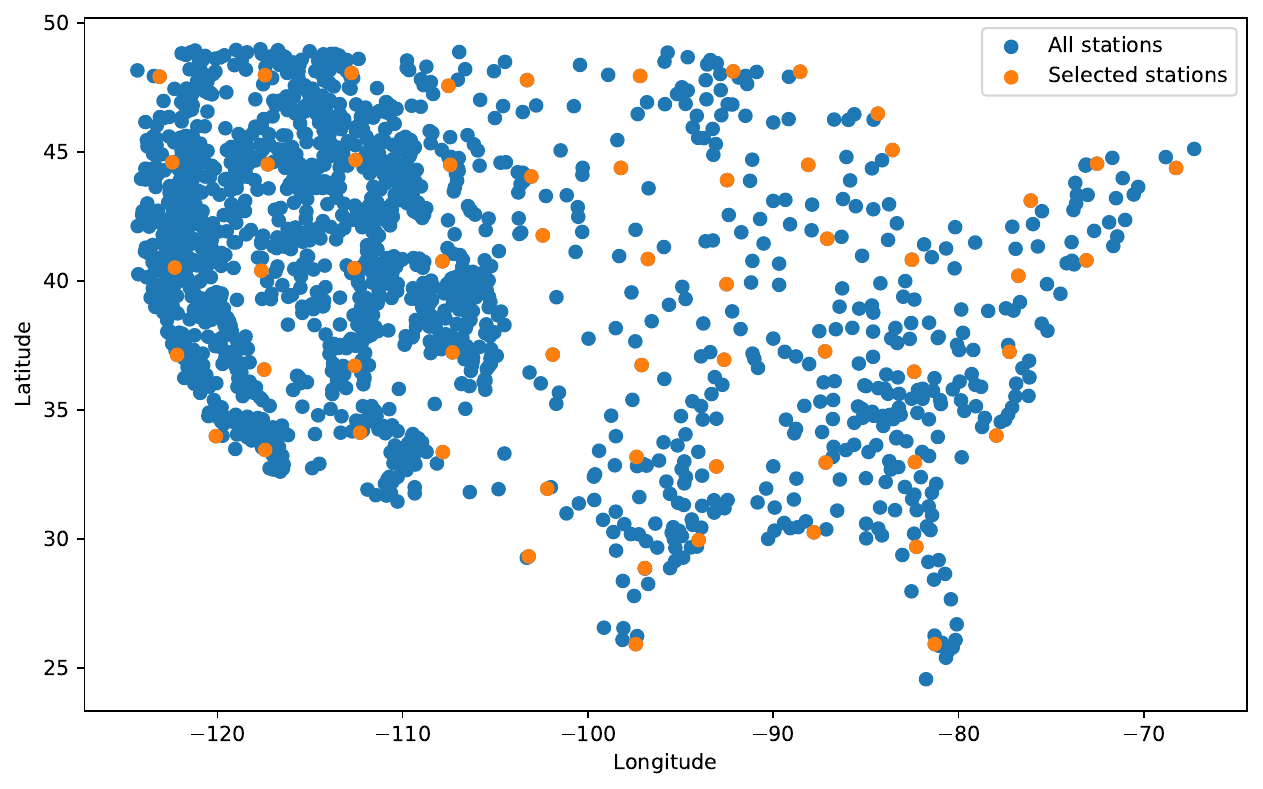}
  \caption{Map of weather stations across the contiguous United States. Orange dots represent observed stations, and blue dots denote unobserved stations used for reconstruction. Observation sites are selected to ensure uniform spatial coverage and mitigate geographic bias.}
  \label{fig:experiments:weather:illustration}
\end{figure}

Among these, 59 stations are selected as observation points, while the remaining stations are treated as unobserved and used for reconstruction. The observation stations are chosen to ensure uniform spatial coverage across the continental U.S. to minimize potential spatial bias. The distribution of observed and unobserved stations is illustrated in Figure~\ref{fig:experiments:weather:illustration}.

\begin{figure}[H]
  \centering
  \begin{subfigure}{0.8\textwidth}
    \includegraphics[width=\textwidth]{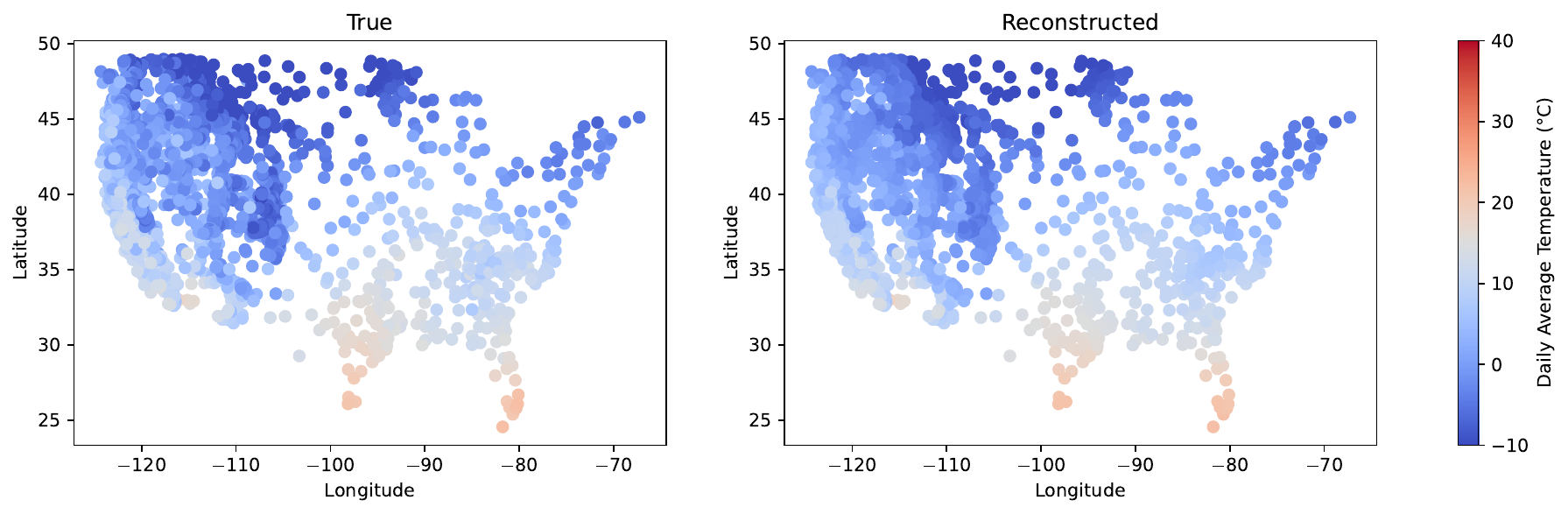}
    \caption{Snapshot on Day 45 of 2024}
  \end{subfigure}
  \begin{subfigure}{0.8\textwidth}
    \includegraphics[width=\textwidth]{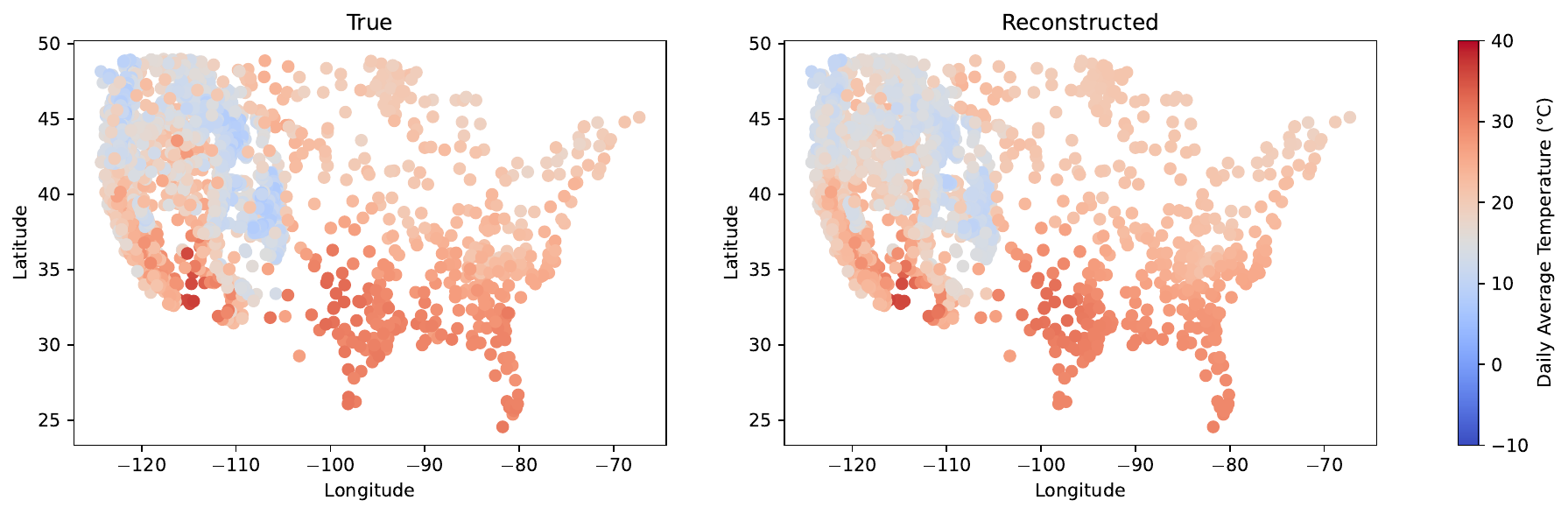}
    \caption{Snapshot on Day 225 of 2024}
  \end{subfigure}
  \caption{True temperature field (left) and reconstructed temperature field (right) on two different days in 2024. The model accurately captures both global temperature trends and local extremes. The reconstruction results closely match the ground truth.}
  \label{fig:experiments:weather:snapshots}
\end{figure}

We use data from 2020 to 2023 for model training and reserve data from 2024 for evaluation. The model is trained for 100 epochs and the training process takes approximately 2 hours on a single NVIDIA A6000 GPU. Reconstruction results for selected winter and summer days in 2024 are shown in Figure~\ref{fig:experiments:weather:snapshots}.

To quantitatively evaluate reconstruction accuracy, we compare our method with a baseline approach that uses the historical average temperature at each station. The historical average is computed as the mean temperature on the same calendar day across the four prior years (2020-2023). Here we use $R^2$ as the metric, which is defined as:
\begin{equation}
    R^2 = 1 - \frac{\sum_{i=1}^{N} (v_i - \hat{v}_i)^2}{\sum_{i=1}^{N} (v_i - \bar{v})^2},
\end{equation}
where $v_i$ is the true value, $\hat{v}_i$ is the predicted value, $\bar{v}$ is the sample mean of the true values, and $N$ is the number of unobserved stations. A higher $R^2$ value indicates better predictive performance. We compare against the same baseline family used for the other cases. As shown in Table~\ref{tab:experiments:weather:baselines}, RFormer obtains the lowest Relative RMSE on the held-out 2024 station reconstructions and the highest mean station-wise $R^2$.

\begin{table}[H]
  \centering
  \begin{tabular}{lcc}
    \toprule
    Method        & Relative RMSE & Station $R^2$ \\
    \midrule
    RFormer       & $\mathbf{0.1757 \pm 0.0862}$ & $\mathbf{0.9465 \pm 0.0664}$ \\
    Gappy POD     & $0.2006 \pm 0.1098$          & $0.9311 \pm 0.0594$          \\
    Voronoi CNN   & $0.2072 \pm 0.0920$          & $0.9246 \pm 0.1047$          \\
    TNP           & $0.2166 \pm 0.0948$          & $0.9179 \pm 0.1164$          \\
    CNP           & $0.3071 \pm 0.1212$          & $0.8228 \pm 0.2568$          \\
    ANP           & $0.3085 \pm 0.1241$          & $0.8206 \pm 0.3018$          \\
    Interpolation & $0.3916 \pm 0.1380$          & $0.6938 \pm 0.4435$          \\
    KRIGING       & $0.4156 \pm 0.1494$          & $0.6607 \pm 0.4904$          \\
    \bottomrule
  \end{tabular}
  \caption{Baseline comparison for the contiguous-U.S. weather case. Relative RMSE is reported as mean $\pm$ standard deviation across snapshots and is lower-is-better. Station $R^2$ is reported as mean $\pm$ standard deviation across stations and is higher-is-better.}
  \label{tab:experiments:weather:baselines}
\end{table}

The weather results show a compact leading group formed by RFormer, Gappy POD, Voronoi CNN, and TNP. RFormer achieves a Relative RMSE of $0.1757$ and station-wise $R^2$ of $0.9465 \pm 0.0664$, followed by Gappy POD ($0.2006$, $R^2=0.9311 \pm 0.0594$), Voronoi CNN ($0.2072$, $R^2=0.9246 \pm 0.1047$), and TNP ($0.2166$, $R^2=0.9179 \pm 0.1164$). CNP and ANP are substantially less accurate, with Relative RMSE near $0.31$, while interpolation and Kriging produce the weakest reconstructions on this sparse station network.

\subsection{3D Blood Flow}
In this experiment, we evaluate our method on the reconstruction of a three-dimensional blood flow dataset simulated using Dissipative Particle Dynamics (DPD). The dataset captures the complex hemodynamics in a realistic vessel geometry, providing a challenging test case for reconstruction due to the irregular domain and highly unstructured data. The goal is to reconstruct the 3D velocity field $(u, v, w)$ from sparse measurements. Notice that the dataset is 3D, but the flow is predominantly in-plane, so we focus on the in-plane velocity components $(u, v)$ for evaluation.

The dataset spans a total of 201 temporal snapshots, where each snapshot contains approximately 300,000 spatial points. We sample 10,000 observation points per snapshot, treating the remaining 290,000 points as query targets to be reconstructed. The model is trained on the first 160 noisy, instantaneous DPD snapshots to learn the flow representation.

A key design choice in this experiment is the train--test protocol: the model is trained on raw, noisy instantaneous snapshots, while evaluation is performed on 20 time-averaged snapshots derived from the later stages of the simulation. This introduces a deliberate distribution shift: the training target is a single noisy realization of the DPD field, whereas the test target is a temporally smoothed mean field that filters out stochastic fluctuations. We adopt this protocol because the raw DPD snapshots are too noisy to serve as a meaningful reconstruction target for comparison. By training on instantaneous data and testing on averaged snapshots, the model is forced to learn the underlying coherent flow structures that persist under time averaging, rather than overfitting to stochastic DPD noise. This setup reflects a practically relevant scenario: inferring physically meaningful, denoised fields from sparse, noisy measurements.

\begin{figure}[H]
  \centering
  \begin{subfigure}{0.8\textwidth}
    \includegraphics[width=\textwidth]{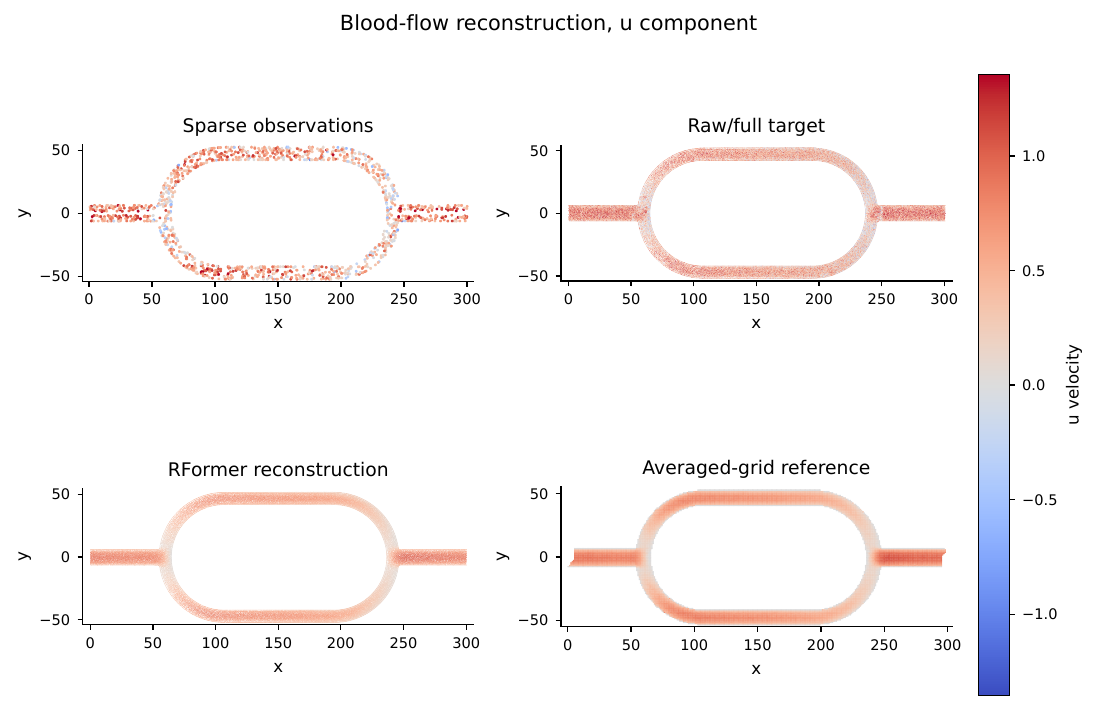}
    \caption{Snapshot of $u$ velocity component}
  \end{subfigure}
  \begin{subfigure}{0.8\textwidth}
    \includegraphics[width=\textwidth]{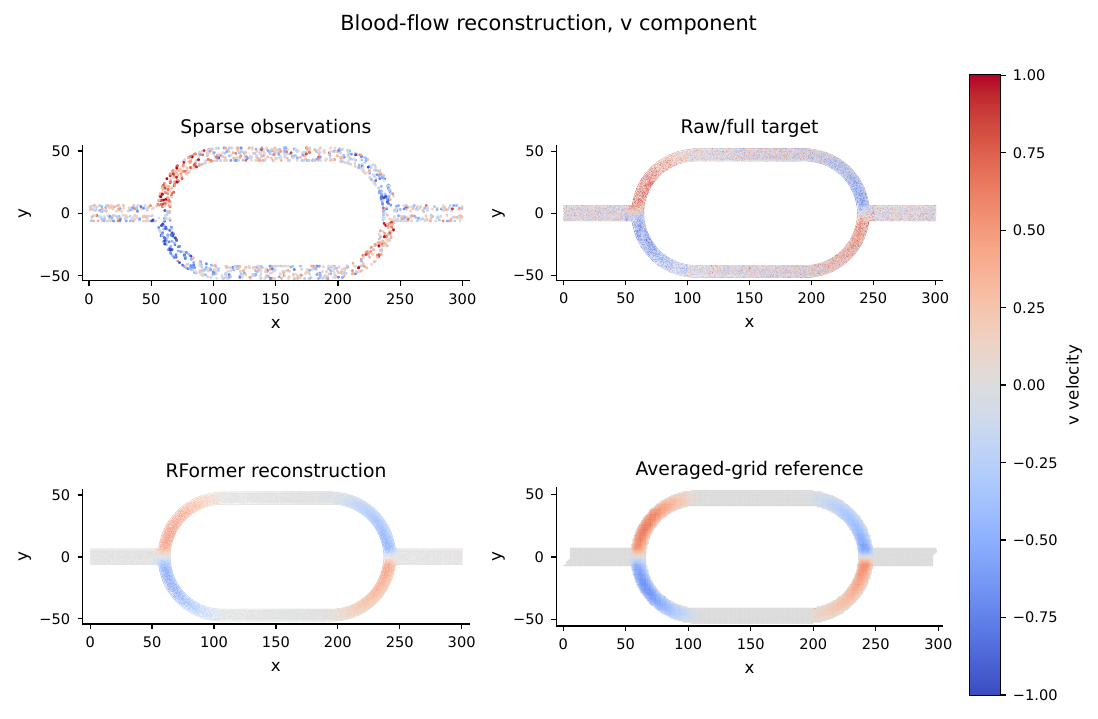}
    \caption{Snapshot of $v$ velocity component}
  \end{subfigure}
  \caption{Reconstruction results for the blood flow dataset at the mid-plane. The figure compares sparse observations, the raw/full target, our reconstruction, and the averaged-grid reference for the $u$ and $v$ velocity components.}
  \label{fig:experiments:blood_flow:snapshots}
\end{figure}

Figure~\ref{fig:experiments:blood_flow:snapshots} presents the mid-plane reconstruction results for the $u$ and $v$ velocity components on a representative validation snapshot. Our method successfully reconstructs the main flow features compared to the averaged-grid reference and the raw target, whereas standard interpolation from sparse observations fails to resolve the fine-scale details.

\begin{figure}[H]
  \centering
  \includegraphics[width=1.0\linewidth]{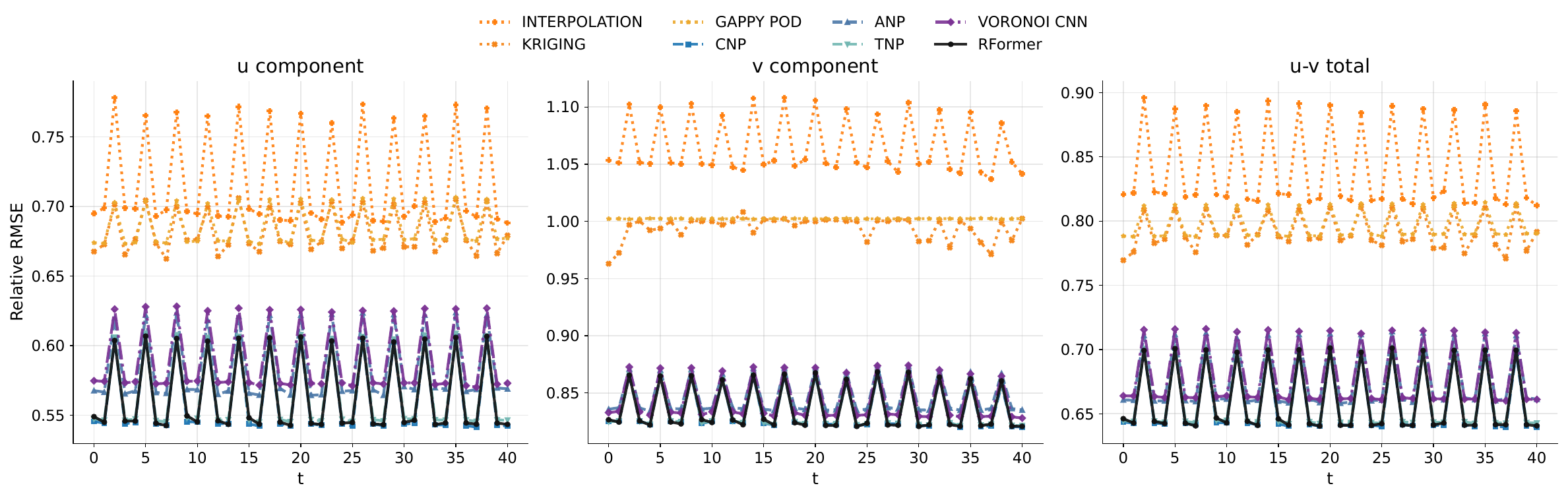}
  \caption{Time-resolved relative RMSE for the $u$, $v$ components and total in-plane velocity field across the validation snapshots. Our method is among the best-performing methods across all snapshots.}
  \label{fig:experiments:blood_flow:error_time}
\end{figure}

To quantify the reconstruction accuracy, we compare our method against interpolation, Kriging, Gappy POD, neural-process baselines, and Voronoi CNN. Since the Voronoi CNN only applies to 2D flow field, we let it use only the x and y coordinates of the data. Table~\ref{tab:experiments:blood_flow:baselines} reports the Relative RMSE for the dominant in-plane velocity components $(u,v)$. RFormer gives the lowest total error, but the margin over TNP and CNP is small, indicating that several context-query neural models perform similarly on this averaged blood-flow target. ANP and Voronoi CNN form the next tier, while Kriging, Gappy POD, and interpolation have noticeably larger errors. As shown in Figure~\ref{fig:experiments:blood_flow:error_time}, our method also reduces the relative RMSE across the validation snapshots compared with the interpolation baselines, and is among the best-performing methods across all snapshots.

\begin{table}[H]
  \centering
  \begin{tabular}{lccc}
    \toprule
    Method         & $u$             & $v$             & Total           \\
    \midrule
    RFormer        & $0.5639 \pm 0.0280$          & $0.8363 \pm 0.0195$          & $\mathbf{0.6604 \pm 0.0267}$ \\
    TNP            & $0.5656 \pm 0.0281$          & $0.8366 \pm 0.0193$          & $0.6616 \pm 0.0267$          \\
    CNP            & $0.5667 \pm 0.0277$          & $0.8389 \pm 0.0182$          & $0.6630 \pm 0.0260$          \\
    ANP            & $0.5846 \pm 0.0255$          & $0.8469 \pm 0.0162$          & $0.6768 \pm 0.0239$          \\
    Voronoi CNN    & $0.5898 \pm 0.0248$          & $0.8437 \pm 0.0185$          & $0.6787 \pm 0.0244$          \\
    KRIGING        & $0.6812 \pm 0.0150$          & $0.9944 \pm 0.0102$          & $0.7916 \pm 0.0135$          \\
    Gappy POD      & $0.6866 \pm 0.0141$          & $1.0051 \pm 0.0007$          & $0.7989 \pm 0.0114$          \\
    Interpolation  & $0.7174 \pm 0.0350$          & $1.0647 \pm 0.0242$          & $0.8405 \pm 0.0332$          \\
    \bottomrule
  \end{tabular}
  \caption{Baseline comparison for the 3D blood-flow case. Values are test-set Relative RMSE reported as mean $\pm$ standard deviation across snapshots for the dominant in-plane velocity components $(u,v)$; lower is better.}
  \label{tab:experiments:blood_flow:baselines}
\end{table}

Quantitatively, RFormer obtains total Relative RMSE $0.6604$, followed closely by TNP at $0.6616$ and CNP at $0.6630$. ANP and Voronoi CNN are moderately higher at $0.6768$ and $0.6787$, respectively. The classical baselines are less effective on this irregular 3D geometry: Kriging reaches $0.7916$, Gappy POD reaches $0.7989$, and interpolation has the largest total error at $0.8405$.

\subsection{3D Turbulent Jet Flow}
Turbulent jet flow experiments are fundamental to understanding the evolution of engine exhaust dynamics, which play a critical role in the design of efficient propulsion systems. Accurate reconstruction of such flows is essential for modeling and controlling jet behavior in both aerospace and industrial applications.

In this example, we evaluate our method on a three-dimensional turbulent jet flow dataset obtained from laboratory experiments conducted at TU Delft\cite{cai2024pinnptv}. The velocity fields were measured using tomographic particle tracking velocimetry (PTV), which provides high-resolution, volumetric flow data. The dataset consists of 200 temporal snapshots, each containing approximately 10,000 spatial points capturing the three velocity components $(u, v, w)$. The flow domain spans a box defined by $[-25, 25] \times [-35, 40] \times [-22, 28]$ in physical units.

\begin{figure}[H]
  \centering
  \includegraphics[width=0.75\linewidth]{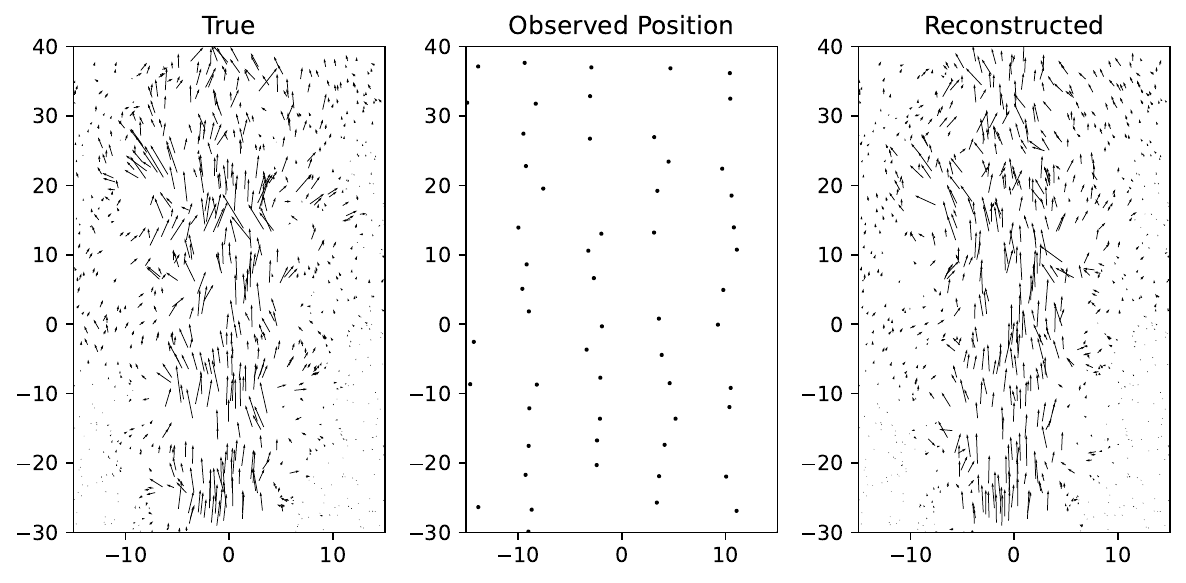}
  \caption{Reconstruction results for snapshot index 180 at the mid-plane $z=3$. Left: true flow field; middle: sampled observation locations; right: reconstructed flow field. The model successfully reconstructs the key spatial features from sparse inputs.}
  \label{fig:experiments:jet_3d:snapshots}
\end{figure}

We randomly sample approximately 1,000 observation points per snapshot to ensure uniform spatial coverage. The remaining points are used for reconstruction. The first 160 snapshots are used for training, while the remaining 40 are held out for evaluation. The model is trained for 100 epochs and the training process takes approximately 14 hours on a single A6000 GPU. A representative reconstruction at $z = 3$ (the midplane of the domain) for snapshot index 180 is shown in Figure~\ref{fig:experiments:jet_3d:snapshots}.

The quantitative baseline comparison is summarized in Table~\ref{tab:experiments:jet_3d:baselines}. RFormer achieves the lowest Relative RMSE for every velocity component and for the total field. Its total error is $0.2540$, compared with $0.4882$ for the next-best method, TNP. CNP and ANP follow with total errors of $0.4971$ and $0.5068$, respectively. Kriging and interpolation are less accurate, with total errors above $0.63$, and Gappy POD has the largest total error at $0.7894$. The largest component-wise gains occur in the streamwise and spanwise components $u$ and $w$, where RFormer roughly halves the error of the neural-process baselines.

\begin{table}[H]
  \centering
  \begin{tabular}{lcccc}
    \toprule
    Method        & $u$             & $v$             & $w$             & Total           \\
    \midrule
    RFormer       & $0.4834 \pm 0.0308$          & $0.1514 \pm 0.0088$          & $0.5185 \pm 0.0271$          & $\mathbf{0.2540 \pm 0.0131}$ \\
    TNP           & $0.9218 \pm 0.0071$          & $0.3197 \pm 0.0092$          & $0.9223 \pm 0.0077$          & $0.4882 \pm 0.0080$          \\
    CNP           & $0.9321 \pm 0.0079$          & $0.3313 \pm 0.0063$          & $0.9275 \pm 0.0066$          & $0.4971 \pm 0.0066$          \\
    ANP           & $0.9533 \pm 0.0047$          & $0.3362 \pm 0.0072$          & $0.9475 \pm 0.0035$          & $0.5068 \pm 0.0067$          \\
    KRIGING       & $1.0027 \pm 0.0208$          & $0.5231 \pm 0.0252$          & $0.9887 \pm 0.0286$          & $0.6360 \pm 0.0175$          \\
    Interpolation & $1.0428 \pm 0.0322$          & $0.5224 \pm 0.0173$          & $1.0748 \pm 0.0428$          & $0.6539 \pm 0.0162$          \\
    Gappy POD     & $1.1301 \pm 0.1631$          & $0.6905 \pm 0.0667$          & $1.1282 \pm 0.1598$          & $0.7894 \pm 0.0823$          \\
    Voronoi CNN   & --                                & --                                & --                                & --                                \\
    \bottomrule
  \end{tabular}
  \caption{Baseline comparison for the 3D turbulent-jet case. Values are test-set Relative RMSE reported as mean $\pm$ standard deviation across snapshots for each velocity component and the total velocity field; lower is better. Voronoi CNN is not available for this unstructured 3D case.}
  \label{tab:experiments:jet_3d:baselines}
\end{table}

\begin{figure}[H]
  \centering
  \includegraphics[width=0.6\linewidth]{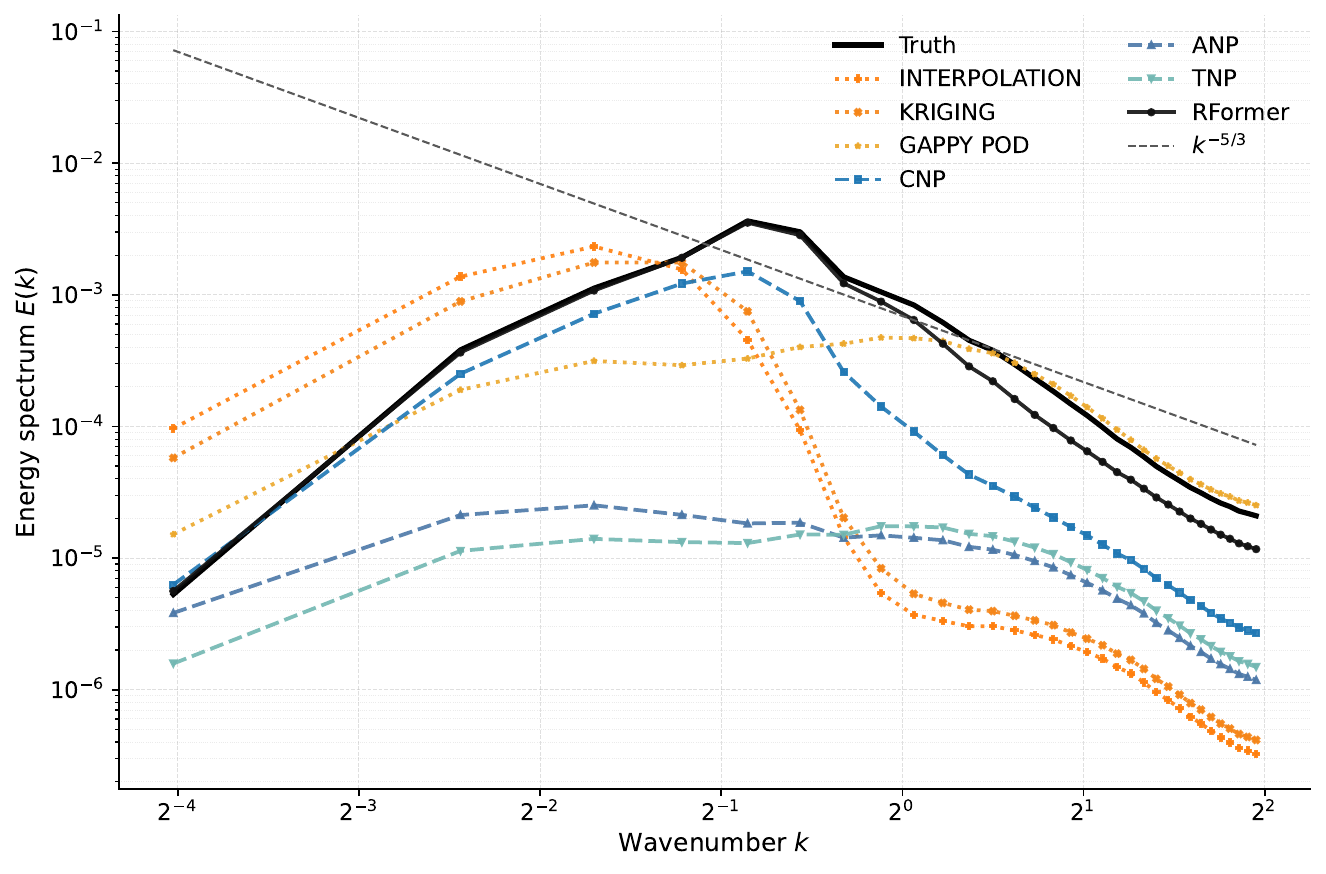}
  \caption{Energy-spectrum comparison for the 3D turbulent-jet case. The x-axis represents spatial wavenumber and the y-axis represents kinetic energy. RFormer closely tracks the ground-truth spectrum over the energetic wavenumber range and preserves the high-wavenumber decay more accurately than the interpolation, Kriging, Gappy POD, CNP, ANP, and TNP baselines. The dashed $k^{-5/3}$ line is shown as a reference scaling.}
  \label{fig:experiments:jet_3d:spectrum}
\end{figure}

For turbulent flows, traditional pointwise error metrics are often insufficient due to the stochastic nature of turbulence. Therefore, we also evaluate reconstruction quality using the energy spectrum, which quantifies the distribution of kinetic energy across spatial scales (wavenumbers). A well-preserved energy spectrum indicates that the reconstructed field captures the flow's multiscale structure, not only its pointwise values. Figure~\ref{fig:experiments:jet_3d:spectrum} compares the energy spectrum of the reconstructed velocity field with the ground truth and the baseline methods. RFormer most closely follows the ground-truth spectral envelope, including the energetic intermediate wavenumbers and the high-wavenumber decay. In contrast, the interpolation and Kriging spectra deviate substantially from the truth, Gappy POD retains only a limited portion of the spectral content, and the neural-process baselines generally underrepresent turbulent energy across much of the spectrum.

\section{Conclusion}

In this study, we introduced a novel operator learning framework that leverages the architecture of language models to reconstruct flow fields from sparse measurements in a mesh-free manner. By casting the reconstruction task as a sequence-to-sequence problem, our approach effectively captures spatial correlations and long-range dependencies inherent in fluid dynamics data.

We validated our method on four different datasets: (1) two-dimensional vortex street simulations, (2) daily average temperature across the contiguous United States, (3) three-dimensional blood flow simulations based on dissipative particle dynamics, and (4) three-dimensional turbulent jet flow measurements acquired via particle tracking velocimetry. In all cases, our model achieved competitive reconstruction accuracy using highly sparse inputs (fewer than 10\% of spatial points observed) and demonstrated computational efficiency at inference time.

The results highlight the potential of integrating language model architectures into operator learning frameworks for scientific and engineering problems. By treating observed data as context and unobserved positions as queries, the model effectively infers missing information like question-answering in natural language processing.

Future work will focus on extending this framework to incorporate temporal dynamics, enabling real-time reconstruction of evolving flow fields with improved accuracy. In addition, we aim to develop foundation models for general flow prediction and reconstruction tasks, which can be fine-tuned for specific applications or used directly via in-context learning. Such advancements could significantly accelerate simulation and design workflows in fluid mechanics and related fields.

\section{Acknowledgements}
The authors from Brown University acknowledge support of the DARPA-APAQuS program grant number  HR00112490526.

\newpage
\bibliographystyle{unsrt}
\bibliography{references}
% Go over citations carefully.
\newpage
\appendix
\section{Baseline Methods and Hyperparameters}
\label{app:baseline_setup}

All baseline experiments use the same data interface as RFormer, with observed coordinates and values stored as $(x,v)$ and query coordinates and values stored as $(qx,qv)$.

\subsection{Classical Baselines}
The interpolation, Kriging, and Gappy POD baselines are fit or solved independently of minibatch training. Interpolation is applied snapshot by snapshot with SciPy's radial-basis-function interpolator using a thin-plate-spline kernel, $128$ nearest neighbors, and smoothing $10^{-8}$; if the RBF solve fails, the implementation falls back to nearest-neighbor interpolation. Kriging is implemented as a Gaussian-process regressor with an RBF kernel plus a white-noise kernel. For each snapshot and output component, at most $512$ observed points are sampled without replacement, using the snapshot index as the random seed. The length scale is $1.0$, the Gaussian-process regularization and white-noise level are both $10^{-6}$, the target is normalized internally, and no optimizer restarts are used. If Gaussian-process fitting fails, the prediction falls back to nearest-neighbor interpolation.

Gappy POD first concatenates each training snapshot's observed and query values into a full reference field on the canonical point set. It computes the temporal mean, performs an SVD of the centered training fields, and keeps the smallest number of POD modes whose cumulative singular-value energy reaches $95\%$. At prediction time, observed sparse values are matched to the POD grid by nearest-neighbor lookup, modal coefficients are recovered by least squares on the observed entries, and the query values are reconstructed from the retained modes.

\subsection{Neural-Process Baselines}
The CNP, ANP, and TNP baselines are trained by minimizing a Gaussian negative log likelihood. The predictive standard deviation is parameterized as
\[
  \sigma = 0.1 + 0.9\,\operatorname{softplus}(\sigma_{\mathrm{raw}}),
\]
in all cases. During training, query points are randomly subsampled when the configuration imposes a limit; during prediction, queries are processed in chunks using the same limit. Table~\ref{tab:appendix:np_baselines} lists the neural-process hyperparameters.

\begin{table}[H]
  \centering
  \resizebox{\textwidth}{!}{%
  \begin{tabular}{lcccccccc}
    \toprule
    Method & Cases & Hidden dim & Layers & Heads & Optimizer & Learning rate & Batch size & Epochs \\
    \midrule
    CNP & all & 128 & encoder 4, decoder 3 & -- & Adam & $10^{-4}$ & 64 & 10000 \\
    TNP & vortex, weather, jet & 128 & transformer 4 & 4 & AdamW & $10^{-4}$ & 64 & 1000 \\
    TNP & blood flow & 128 & transformer 4 & 8 & AdamW & $10^{-4}$ & 16 & 1000 \\
    ANP & vortex, weather, jet & 128 & encoder/decoder 4 & 4 & AdamW & $10^{-4}$ & 64 & 1000 \\
    ANP & blood flow & 128 & encoder/decoder 4 & 8 & AdamW & $10^{-4}$ & 16 & 1000 \\
    \bottomrule
  \end{tabular}
  }
  \caption{Neural-process baseline hyperparameters. CNP uses an MLP encoder with mean aggregation and an MLP decoder. TNP uses a transformer encoder with a learned context/query role embedding and a two-layer prediction head. ANP uses deterministic cross-attention, a latent encoder, and a KL weight of $1.0$. Dropout is $0$ for ANP and TNP.}
  \label{tab:appendix:np_baselines}
\end{table}

For the vortex-street, weather, and turbulent-jet cases, CNP, ANP, and TNP include the context points as supervised targets and use at most $2048$ query points per training batch, with no context subsampling. For the blood-flow case, ANP and TNP use at most $1024$ context points and $1024$ query points per batch and do not include context points as targets; CNP keeps the common configuration used in the other cases.

\subsection{Voronoi CNN Baseline}
The Voronoi CNN baseline rasterizes the sparse observations before applying a convolutional network. For each snapshot, observed values are assigned to an automatically inferred regular grid by nearest-neighbor Voronoi rasterization in the selected two coordinate axes. A binary observation mask is concatenated to the rasterized value channels. The CNN has seven hidden convolutional layers with $48$ channels, ReLU activations, and kernel size $7$. The predicted dense raster is sampled back at query coordinates by bilinear grid sampling. Values are normalized by the training-set mean and standard deviation. Training uses Adam with learning rate $10^{-3}$ and early stopping with patience $100$ based on validation MSE.

\begin{table}[H]
  \centering
  \begin{tabular}{lcccc}
    \toprule
    Case & Batch size & Max epochs & Final kernel & Coordinate axes \\
    \midrule
    Vortex street & 128 & 5000 & 3 & $(0,1)$ \\
    Weather & 32 & 5000 & 7 & $(0,1)$ \\
    Blood flow & 32 & 250 & 7 & $(0,1)$ \\
    Turbulent jet & - & - & - & - \\
    \bottomrule
  \end{tabular}
  \caption{Case-specific Voronoi CNN settings. All cases use an automatically inferred grid size, value normalization, seven $7\times 7$ hidden convolutions with $48$ channels, and validation-based early stopping. For blood flow dataset, the Voronoi rasterization is performed on the first two coordinate axes. It is not applied to the turbulent-jet case because the turbulent jet dataset is essentially 3D.}
  \label{tab:appendix:voronoi_cnn}
\end{table}

\end{document}